\documentclass[prl,superscriptaddress,showpacs,preprint]{revtex4}
\usepackage{amsmath,graphicx,amssymb,amsfonts}
\begin{document}\title{From Elasticity to Hypoplasticity:
Dynamics of Granular Solids}
\author{Yimin Jiang}
\affiliation{Theoretische Physik, Universit\"{a}t T\"{u}bingen,72076 T\"{u}bingen, Germany}
\affiliation{Central South University, Changsha 410083, China}
\author{Mario Liu}
\affiliation{Theoretische Physik, Universit\"{a}t T\"{u}bingen,72076 T\"{u}bingen, Germany}
\date{\today}
\begin{abstract}
``Granular elasticity," useful for calculating static
stress distributions in granular media, is generalized by
including the effects of slowly moving, deformed grains.
The result is a hydrodynamic theory for granular solids
that agrees well with models from soil mechanics.
\end{abstract}
\pacs{81.40.Lm, 83.60.La, 46.05.+b, 45.70.Mg} \maketitle

Granular media has different phases that, in dependence
of the grain's ratio of deformation to kinetic energy,
may loosely be referred to as gaseous, liquid and solid.
The first phase is relatively well understood: Moving
fast and being free most of the time, the grains in the
gaseous phase have much kinetic, but next to none
elastic, energy~\cite{Haff}. In the denser liquid phase,
say in chute flows, there is less kinetic energy, more
deformation, and a rich rheology that has been
scrutinized recently~\cite{chute}. In granular statics,
with the grains deformed but stationary, the energy is
all elastic. This state is legitimately referred to as
solid because static shear stresses are sustained. If
granular solid is slowly sheared, the predominant part of
the energy remains elastic. Yet no theory is capable of
accounting for both its statics and dynamics, and no
picture exists that helps to render its physics
transparent.

Two grains in contact are initially very compliant,
because so little material is being deformed. As this
geometric fact should also hold on larger scales, for
many grains,  {\em diverging compliance at diminishing
compression} is a basic characteristics of granular
solids, and the reason it is sensible to abandon the
approximation of infinitely rigid grains. Starting from
this observation, a theory termed {\sc ge} (for
``granular elasticity") was constructed to account for
static granular stress distributions. Taking the energy
$w$ as a function of $u_{ij}$, the elastic contribution
to the total strain field $\varepsilon_{ij}$, we
specify~\cite{J-L}
\begin{equation}
w=\sqrt{\Delta }\left( {\mathcal B}\textstyle\frac25
\Delta ^2+  {\mathcal A}u_s^2\right)={\mathcal
B}\sqrt{\Delta }\left(\textstyle\frac25\Delta ^2+
u_s^2/\xi \right), \label{1}
\end{equation}
with $\Delta\equiv -u_{\ell\ell}$, $u_s^2\equiv
u^0_{ij}u^0_{ij}$, $u^0_{ij}\equiv u_{ij}-\frac13
u_{\ell\ell}\,\delta_{ij}$. (The notations:
$a^0_{ij}\equiv a_{ij}-\frac13 a_{\ell\ell}\,\delta_{ij}$
and $a_s^2\equiv a^0_{ij}a^0_{ij}$ with any $a_{ij}$ are
employed throughout this paper.) The elastic coefficient
$\mathcal{B}$, a measure of overall rigidity, is a
function of the density. Denoting $\rho_g$ as the
granular material's bulk density, and
$e\equiv\rho_g/\rho-1$ as the void ratio, we take
$\mathcal{B}= \mathcal{B}_0\times{(2.17-e) ^2}/[1.3736(
1+e)]$, with $\mathcal{B}_0,\xi>0$ two material
constants. The elastic energy $w$ contributes
$\pi_{ij}\equiv-\partial w/\partial u_{ij}$ to the total
stress $\sigma_{ij}$. And since the elastic stress is the
only contribution in statics, force balance reads
$\nabla_j\sigma_{ij}=\nabla_j\pi_{ij}=\rho G_i$. This was
solved for three classical cases: silos, sand piles and
granular sheets under a point load, resulting in rather
satisfactory agreement to experiments, see~\cite{ge}.
Moreover, the energy $w$ (with $P\equiv\frac13\pi _{\ell
\ell }$) is convex only for $\pi _s/P\le\sqrt{2/\xi}$,
implying no elastic solution is stable beyond it.
Identifying this as the yield surface gives
$\xi\approx5/3$ for natural sand.

When granular solid is being slowly sheared, we must
expect a qualitative change of its behavior: In addition
to moving with the large-scaled velocity $v_i$, the
grains also move and slip in deviation of it -- implying
a small but finite granular temperature $T_g$. As a
result, some of the grains are temporarily unjammed, with
enough time to decrease their deformation. This depletes
the elastic energy and relaxes the static stress. Stress
relaxation is typical of viscoelastic systems such as
polymers. Granular media are similar, but they possess a
relaxation rate that vanishes with $T_g$. This is the
reason they return to perfect elasticity when stationary.
The basic physics of granular solids, {\em
viscoelasticity at finite} $T_g$, is in fact epitomized
by a sand pile, which holds its shape when unperturbed,
but fails to do so when tapped. A set of differential
equations termed {\em granular solid hydrodyna\-mics}
({\sc gsh}) is derived consistently below starting from
{\sc ge}, with this simple physics as the only additional
input.

Conservation of density and momentum always holds,
\begin{equation}
{\textstyle\frac\partial{\partial t}}\rho+\nabla _i(\rho
v_i)=0,\ \ \ {\textstyle\frac\partial{\partial
t}}(\rho{v}_i)+\nabla _j(\sigma_{ij}+\rho v_iv_j)=\rho
G_i, \label{2}
\end{equation}
where $G_i$ is the gravitational constant. In granular
gas or liquid, the stress $\sigma_{ij}$ has the same
structure as in the Navier-Stokes equation, though the
viscosity is a function of the shear. In granular solid,
the stress is not usually taken to be given in a closed
form. Instead, constitutive relations are employed. These
relate the temporal derivatives of stress and strain,
giving ${\textstyle\frac\partial{\partial t}}\sigma_{ij}$
as a function of $v_{ij}\equiv\frac12(\nabla _iv_j+\nabla
_jv_i)$ and density (where
${\textstyle\frac\partial{\partial t}}$ is often replaced
by an objective derivative say from Jaumann).

Hypoplasticity, or {\sc hpm} (for hypoplastic model), is
a modern, well-verified, yet comparatively simple theory
of soil mechanics~\cite{Kolym}. It is quite realistic in
the above specified regime of solid dynamics, though less
appropriate for determining static stress distributions.
The starting point is the rate-independent constitutive
relation,
\begin{equation}
{\textstyle\frac\partial{\partial t}}{\sigma }_{ij}
=H_{ijk\ell}v_{k\ell}+\Lambda _{ij}\sqrt{v_{\ell
k}^0v_{\ell k}^0+\epsilon \left( v_{\ell\ell}\right) ^2},
\label{hp1}
\end{equation}
where the coefficients $H_{ijk\ell},\Lambda
_{ij},\epsilon$ are functions of $\sigma_{ij},\rho$,
specified using experimental data mainly from triaxial
apparatus. Great efforts are invested in finding accurate
expressions for them, of which a recent set~\cite{Kolym}
is $\epsilon=1/3$,
\begin{eqnarray}\label{hp2}
H_{ijk\ell} &=&f\left( F^2\delta _{ik}\delta
_{j\ell}+a^2\sigma _{ij}\sigma _{k\ell}/\sigma
_{nn}^2\right) \text{,}
\\ \Lambda _{ij} &=&aff_dF\left( \sigma
_{ij}+\sigma _{ij}^0\right) /\sigma _{nn}, \label{hp3}
\end{eqnarray}
where  [with $a=2.76$, $h_s=1600$~MPa,$\ e_d=0.44e_i$,
$e_c=0.85e_i$, $e_i^{-1}=\exp \left( \sigma
_{\ell\ell}/h_s\right) ^{0.19}$, $e$ the void ratio]
\begin{eqnarray*} f_d =\left(
\frac{e-e_d}{e_c-e_d}\right) ^{0.25}\!\!\!\!\!\!,\quad
f=-\frac{ 8.7h_s\left( 1+e_i\right) }{3\left(
\sigma_s/\sigma_{\ell\ell}+1\right) e}\left( \frac{\sigma
_{\ell\ell}}{ h_s}\right) ^{0.81}\!\!\!\!\!\!,
\\ \nonumber F =\sqrt{\frac{3\sigma _s^2}{8\sigma _{\ell\ell}
^2}+\frac{2\sigma _s^2\sigma _{\ell\ell}-3\sigma
_s^4/\sigma _{\ell\ell}}{2\sigma _s^2\sigma
_{\ell\ell}-6\sigma _{ij}^0\sigma _{j\ell}^0\sigma _{\ell
i}^0}}-\sqrt{\frac 38}\frac{\sigma _s}{\sigma
_{\ell\ell}}.
\end{eqnarray*}

If {\sc gsh} as derived below from the idea given above
reduces to {\sc hpm} under certain conditions, we would
have, on one hand, captured valuable insights into the
physics of this field-tested theory, understood its range
of validity,  how to widen it by appropriate
modifications, and on the other hand, obtained a
broadside verification of {\sc gsh}, along with the
physical picture embedded in it. As we shall see, {\sc
gsh} indeed reduces to Eq~(\ref{hp1}) for a stationary
$T_g$, with $H_{ijlk},\Lambda _{ij},\epsilon$ given in
terms of $M_{ijk\ell}\equiv-\partial^2 w/\partial
u_{ij}\partial u_{k\ell}$ (known from {\sc ge}) and four
new scalars [combinations of transport coefficients such
as viscosities and stress relaxation rates, see
Eq~(\ref{last})]. Although quite different from
Eqs~(\ref{hp2},\ref{hp3}), the new $H_{ijlk},\Lambda
_{ij},\epsilon$ yield very similar accounts in all cases
we have considered.

A large part of  {\sc gsh} may be duplicated from the
hydrodynamic theory of transient elasticity,
constructed to describe polymers~\cite{temmen}. This
theory accounts for any system in which both the
elastic energy and stress relax, irrespective how this
happens microscopically -- whether due to polymer
strands disentangling, or the grains unjamming. (A
formal and rather more detailed derivation of  {\sc
gsh} can be found in an accompanying paper~\cite{7}.)
The stress $\sigma_{ij}$ and the elastic strain
$u_{ij}$ are determined  by
\begin{equation}
\label{gsh1} \sigma_{ij}=\pi_{ij}-\sigma^D_{ij},\qquad
({\textstyle\frac\partial{\partial t}}+v_k\nabla_k)\,
u_{ij} =v_{ij}+X_{ij},
\end{equation}
where $\pi_{ij}\equiv-\partial w/\partial u_{ij}$ is the
elastic stress and $v_{ij}\equiv\frac12(\nabla
_iv_j+\nabla _jv_i)$. $\sigma^D_{ij}$ and $X_{ij}$ are
the irreversible contributions, given by Onsager
relations that connect the ``currents," $\sigma^D_{ij},
X_{ij}$, to the ``forces," $v_{ij}, \pi_{ij}$,
\begin{eqnarray}\label{gsh2}\sigma^D_{ij}=(\eta+\eta_g)v^0_{ij}
+(\zeta+\zeta_g)\delta_{ij}v_{\ell\ell}+\alpha\pi_{ij},
\\
X_{ij}=-\alpha v_{ij}+\beta\pi^0_{ij}+\beta_1
\delta_{ij}\pi_{\ell\ell}\,\,\\ =-\alpha
v_{ij}-{\textstyle\frac1\tau} u^0_{ij}
-{\textstyle\frac1{\tau_1}}\delta_{ij}u_{\ell\ell}.\label{gsh3}
\end{eqnarray}
The coefficients $\eta,\zeta,\eta_g,\zeta_g>0$ in
$\sigma^D_{ij}$ are viscosities, see below for their
differences. Calculating ${\textstyle\frac\partial
{\partial t}}{\sigma }_{ij}$ as in Eq~(\ref{hp1}),
they all vanish for steady velocities,
${\textstyle\frac\partial{\partial t}} {v}_{i}=0$. The
term $X_{ij}$, accounting for the relaxation of the
elastic strain $u_{ij}$, is rather more consequential.
Eq~(\ref{gsh3}) is obtained by taking the derivative
of Eq~(\ref{1}), $\pi_{ij}\equiv-\partial w/\partial
u_{ij} =\sqrt\Delta({\cal B}\Delta\,\delta
_{ij}-2{\cal A}\, u_{ij}^0) +{\cal A}
({u_s^2}/{2\sqrt\Delta})\delta _{ij}$. So the
relaxation times are given as $1/\tau\equiv2\beta{\cal
A}\sqrt\Delta$,
$1/\tau_1\equiv3\beta_1\sqrt\Delta({\cal
B}+{\textstyle\frac12}{\cal A}u_s^2/\Delta^2)$. The
coefficient $\alpha$ is a cross coefficient of the
Onsager matrix. It is taken as a scalar  for
simplicity.

In principle, the transport coefficients  $\eta$,
$\eta_g$, $\zeta$, $\zeta_g$, $\tau$, $\tau_1$,
$\alpha$ are functions of the thermodynamic variables:
density, temperature and the elastic strain $u_{ij}$.
We shall, again for simplicity, assume that they are
strain-independent, while noting three points:
(1)~Constant $\tau,\tau_1$ implies strain-dependent
$\beta,\beta_1$. Choosing the former as constant and
not the latter, the trace and traceless part of
${\textstyle\frac\partial{\partial t}}u_{ij}$ are
decoupled. (2)~As discussed above, $1/\tau,1/\tau_1$
vanish with $T_g$. So the obvious and simplest
assumption is
\begin{equation}\label{tau}
1/\tau=\lambda T_g,\quad 1/\tau_1=\lambda_1 T_g,
\end{equation}
with $\lambda,\lambda_1, \tau_1/\tau=\lambda/\lambda_1$
possibly functions of the density, but independent from
stress and $T_g$.  (3)~Being reactive, $\alpha$ is not
restricted in its magnitude. It may stay constant while
$1/\tau, 1/\tau_1$ vary -- though it must eventually
vanish for $1/\tau,1/\tau_1\to0$, as $\alpha=0$ in
statics.

The above hydrodynamic theory is closed if we amend it
with an equation of motion for $T_g$. In
thermodynamics, the energy change ${\rm d}w$ from all
microscopic, implicit variables is subsumed as $T{\rm
d}s$, with $s$ the entropy and $T\equiv\partial
w/\partial s$ its conjugate variable. From this, we
divide out the kinetic energy of granular random
motion, executed by the grains in deviation from the
ordered, large-scale motion, and denote it as $T_g{\rm
d}s_g$, calling $s_g,\, T_g\equiv\partial w/\partial
s_g$ granular entropy and temperature. In other words,
we consider two heat reservoirs, the first containing
the energy of granular random motion, the second the
rest of all microscopic degrees of freedom, especially
phonons. In equilibrium, $T_g=T$, and $s_g$ is part of
$s$. But when the granular system is being tapped or
sheared, and $T_g$ is many orders of magnitude larger
than $T$, then this leaky, intermediary heat reservoir
produces physics in its own right. Taking $s_g$ as the
part of the entropy accounting for the granular
kinetic energy, our definition is fairly close to the
entropy of granular gas~\cite{Haff}, though its
functional dependence is probably dominated by the
effect of excluded volumes. The entropy $s$, on the
other hand, is closer to the so-called
``configurational entropy,"~\cite{Edw} (see section 6
of the first of~\cite{ge} for a discussion of their
relationship). The balance equations are
${\textstyle\frac\partial {\partial t}}s
+\nabla_k(sv_k)=R/T$, ${\textstyle\frac\partial
{\partial t}}s_g+ \nabla_k(s_gv_k)=R_g/T_g$, where
\begin{eqnarray}\label{ep2}
R&=&\eta v_s^2+\zeta
v_{\ell\ell}^2+\beta\pi_s^2+\beta_1\pi_{\ell\ell}^2+\gamma
T_g^2,\\ R_g&=&\eta_g v_s^2+\zeta_g v_{\ell\ell}^2-\gamma
T_g^2. \label{ep1}
\end{eqnarray}
The first four terms in the entropy production $R$ are
the usual contributions from shear flow and stress
relaxation, as given by transient elasticity. The
first two terms of $R_g$ account analogously for shear
excitation of random motion. The term $\gamma T_g^2$
(with $\gamma>0$) describes how the kinetic energy of
random motion seeps from $s_g$ into $s$. (Diffusion of
$T,T_g$ are easily included when needed.)

With Eqs~(\ref{1}, \ref{2}, \ref{gsh1}, \ref{gsh2},
\ref{gsh3}, \ref{tau}, \ref{ep1}, \ref{ep2}), {\sc gsh}
is complete. It especially contains the equilibrium case,
$\sigma_{ij}=\pi_{ij}$, in which the dissipative fields
vanish, $\sigma^D_{ij},X_{ij}=0$. Off equilibrium, these
two fields are finite, and we calculate
${\textstyle\frac\partial {\partial t}}\sigma_{ij}$
assuming ${\textstyle\frac\partial{\partial t}}
{v}_{i}=0$, from Eqs~(\ref{gsh1}, \ref{gsh2},
\ref{gsh3}),
\begin{eqnarray}\nonumber
{\textstyle\frac\partial {\partial
t}}\sigma_{ij}=(1-\alpha){\textstyle\frac\partial
{\partial t}}\pi_{ij}=(1-\alpha)M_{ijk\ell}
{\textstyle\frac\partial {\partial t}}u_{k\ell}=\\
{\textstyle(1-\alpha) M_{ijk\ell}
[(1-\alpha)v_{k\ell}-\frac1\tau
u^0_{k\ell}-\frac1{\tau_1}\delta_{k\ell}u_{\ell\ell}]}.
\label{gsh4}\end{eqnarray}

As mentioned above, the energy $w$ looses its convexity
at $\pi_s/P=\sqrt{6/5}$, and no static, elastic solution
is possible beyond this ratio. Therefore, it was
identified as yield. Given Eq~(\ref{gsh4}), the same
identification holds dynamically: The loss of convexity
implies that one of the six eigenvalues of
$M_{ijk\ell}\equiv-\partial^2w/\partial u_{ij}\partial
u_{k\ell}$ (written as a $6\times6$ matrix) vanishes at
this point, and a strain rate along the associated
direction yields vanishing stress rate.

For $R_g=0$, when $s_g$ is being produced and leaking
at the same rate, we have a stationary $T_g$, given as
\begin{equation}\label{tg}
T_g={\sqrt{\eta_g/\gamma}\sqrt{v_s^2+(\zeta _g/\eta
_g)v_{\ell\ell}^2}}.
\end{equation}
Inserting Eqs~(\ref{tau},\ref{tg}) into (\ref{gsh4}),
we retrieve Eq~(\ref{hp1}), with
\begin{eqnarray}\label{hpA}
H_{ijk\ell}=(1-\alpha)^2 M_{ijk\ell},\qquad
\epsilon=\zeta _g/\eta _g,\\
\Lambda_{ij}=(1-\alpha)M_{ijk\ell}
[(\tau/\tau_1)\Delta\delta_{k\ell}-u_{k\ell}^0]\lambda
{\sqrt{\eta_g/\gamma}}. \label{hpB}\end{eqnarray}
{\sc hpm} has 43 free parameters (36+6+1 for
$H_{ijk\ell}, \Lambda_{ij}, \epsilon$), all functions of
the stress and density. Expressed as here, the stress and
density dependence are essentially determined by
$M_{ijk\ell}$ that (with $\xi=5/3$ and ${\cal
B}_0=8500$~MPa) is a known quantity~\cite{ge}. For the
four free constants, we take
\begin{equation}\label{last}
1-\alpha=0.22,\,\,\frac\tau{\tau_1}=0.09,\,\,\frac{\zeta
_g}{\eta _g}=0.33,\,\,\lambda
\sqrt{\frac{\eta_g}{\gamma}}=114,
\end{equation}
to be realistic choices, as these numbers yield
satisfactory agreement with {\sc hpm}. Their significance
are: $\zeta _g/\eta _g=0.33$ implies shear flows are
three times as effective in creating $T_g$ as
compressional flows. $\tau/\tau_1=0.09$ means, plausibly,
that the relaxation rate of shear stress is ten times
higher than that of pressure. For a purely elastic
system, Eq~(\ref{hp1}) is replaced by ${\textstyle\frac
\partial{\partial t}} {\sigma }_{ij} =M_{ij\ell k}v_{\ell
k}$. Therefore, the factor $(1-\alpha)^2$ accounts for an
overall, dynamic softening of the static compliance
tensor $M_{ij\ell k}$, a known effect in soil
mechanics~\cite{her}. Finally, $\lambda$ controls the
stress relaxation rate for given $T_g$, and
${\textstyle\sqrt{\eta_g/\gamma}}$ how well shear flow
excites $T_g$. Together, $\lambda{\textstyle\sqrt
{\eta_g/\gamma}}=114$ determines the relative weight of
plastic versus reactive response. (Note
$|\Lambda_{ij}|/|H_{ijk\ell}|
\sim|u_{k\ell}^0|\cdot114/(1-\alpha)$ is, for
$|u_{ij}^0|$ around $10^{-3}$, of order unity.)

\begin{figure}[t]
\begin{center}
\includegraphics[scale=0.8]{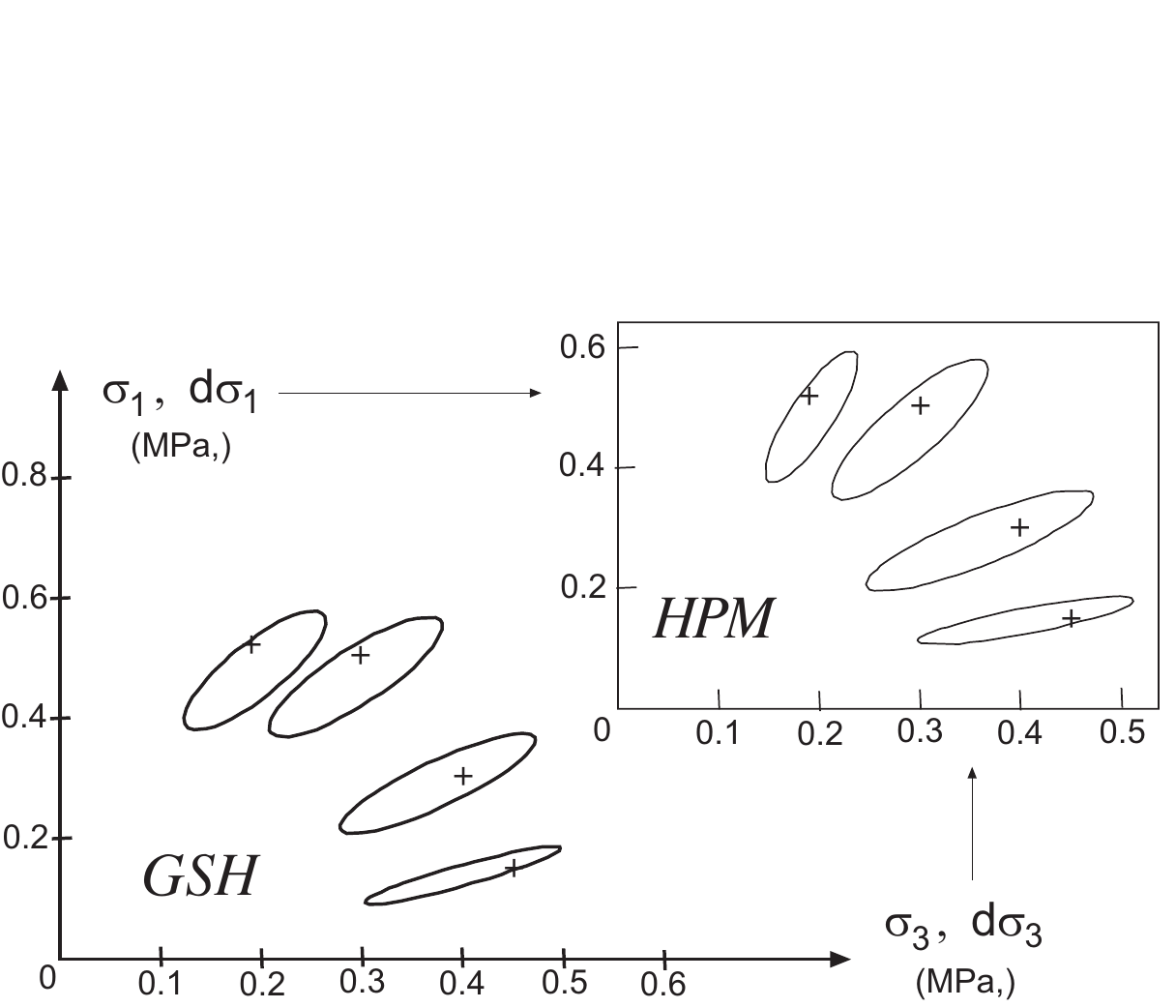}
\end{center}
\caption{The stress changes ${\rm d}\sigma_1, {\rm
d}\sigma_3$, calculated using {\sc gsh} (granular solid
hydrodynamics) and {\sc hpm} (hypoplastic model), for
given strain rate starting from different points
(depicted as crosses) in the stress space spanned by
$\sigma_1, \sigma_3$. The strain rate has varying
directions but a constant amplitude,
$\sqrt{2v_1^2+v_3^2}$, such that the applied strain
changes form circles around each cross (not shown).}
\label{fig1}
\end{figure}
\begin{figure}[b]
\begin{center}
\includegraphics[scale=0.85]{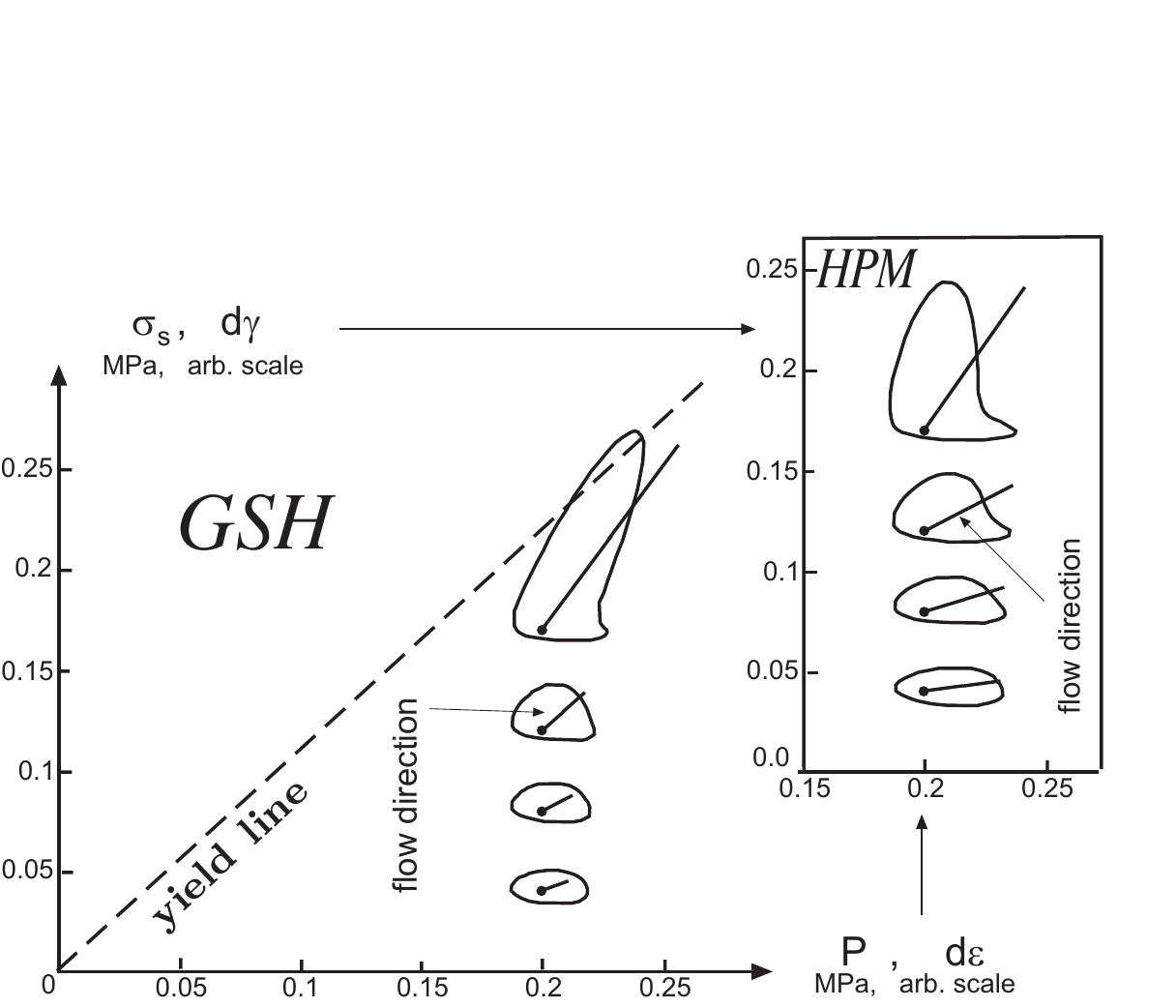}
\end{center}
\caption{The change in strain ${\rm d}\gamma,{\rm
d}\varepsilon$ for given stress rate starting from
different points in the stress space, spanned by
$\sigma_s,P$. The amplitude of the stress rate
$\sqrt{{\rm d}P^2+{\rm d}q^2}$ is constant. See Fig 3 for
an explanation of the ``flow direction."} \label{fig1a}
\end{figure}
Next, we compare Eqs~(\ref{hpA}, \ref{hpB}) to
(\ref{hp2}, \ref{hp3}) in their results with respect to
``response envelopes," a standard test in soil mechanics
for rating constitutive relations~\cite{Kolym}. Axial
symmetry of the triaxial geometry is assumed, with
$\sigma_{ij}, v_{ij}$ diagonal, and
$\sigma_1\equiv\sigma_{xx}=\sigma_{yy}$,
$\sigma_3\equiv\sigma_{zz}$, $v_1\equiv v_{xx}=v_{yy}$,
$v_3\equiv v_{zz}$,
$P\equiv\frac23\sigma_1+\frac13\sigma_3$,
$q\equiv\sigma_3-\sigma_1$,
$\sigma_s^2\equiv{\frac23}q^2$, ${\rm d}\gamma\equiv
(v_1-v_3){\rm d}t$, ${\rm
d}\varepsilon\equiv-(2v_1+v_3){\rm d}t$. Starting from a
point in the stress space (spanned by $\sigma_1,\sigma_3$
in Fig 1 and $\sigma_s,P$ in Fig~2), one deforms the
system for a constant time ${\rm d}t$, at given strain or
stress rates, while recording the change in the conjugate
quantity. Varying the direction, the input is a circle
around the starting point, but the response envelopes
show deformation characteristic of the system, or the
constitutive relation to be rated. Fig~1 and 2 show
respectively the responding stress and strain envelopes,
for the void ratio $e=0.66$, calculated using {\sc gsh}
and {\sc hpm}. The similarity in stress-dependence and
anisotropy is obvious.

\begin{figure}[t]
\begin{center}
\includegraphics[scale=0.85]{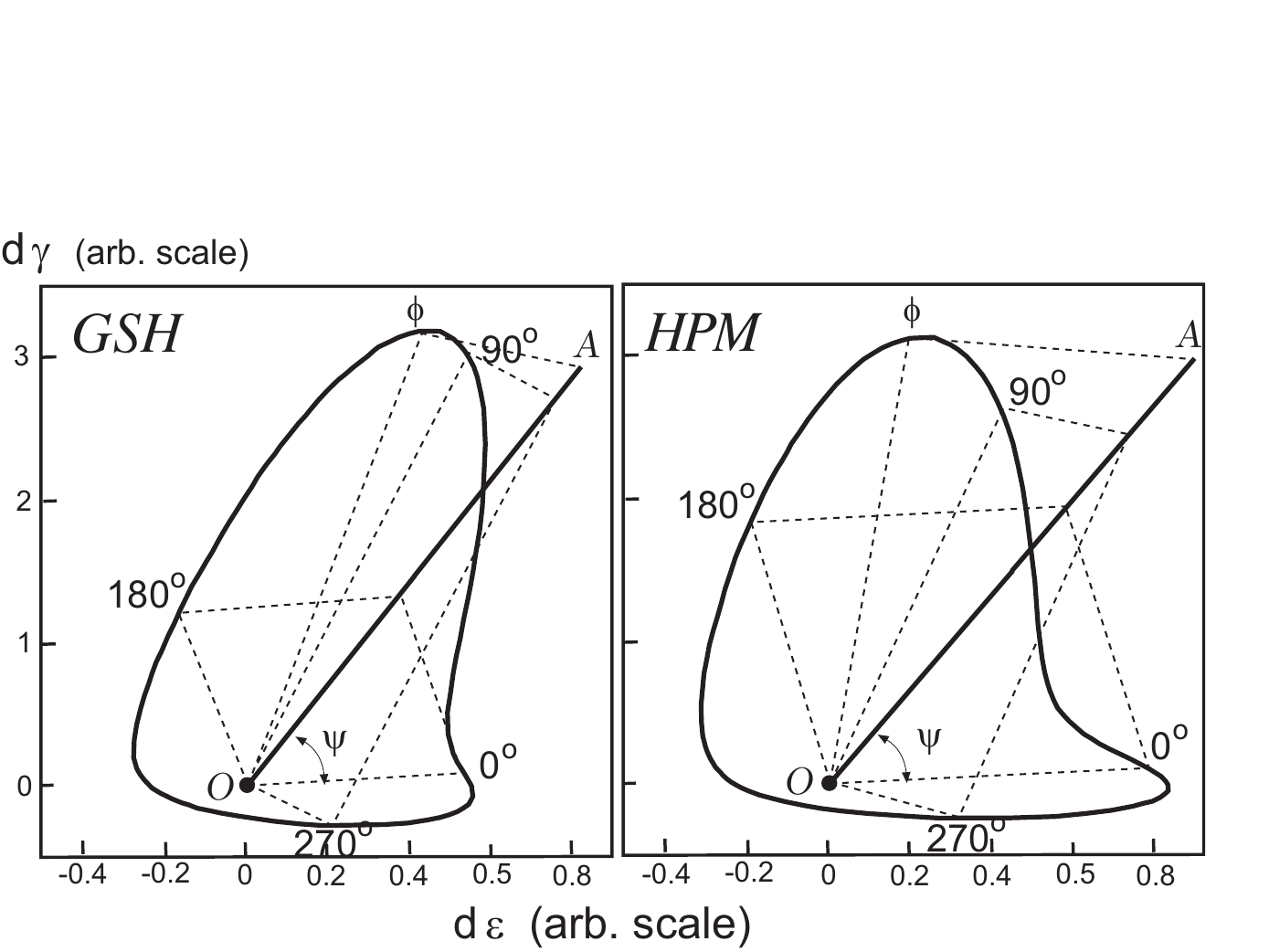}
\end{center}
\caption{A pair of blown-up strain envelopes from Fig 2,
with the starting point O at $P=0.2$, $\sigma_s=0.16$
MPa. The stress rate is reversed at halftime, and the
stress returns to the origin O at the end. The strain
(depicted as dotted lines) gets deflected, and ends
somewhere along OA, a straight line for both {\sc gsh}
and {\sc hpm}. $\sigma$, the angle of OA, is called the
``flow direction;" $\phi$ is the ``yield direction,"
along which the plastic flow is maximal, with the strain
ending at A.} \label{fig2}
\end{figure}

In Fig 3, one strain envelope is blown up for a more
detailed comparison, using the extended version of
response envelope as given in~\cite{hh}. Here, the
applied stress rate is reversed at halftime, such that
the system returns to the starting point in stress space
at the end. The responding strain change, depicted as
deflected, straight dotted lines, does not return to the
origin. Both {\sc gsh} and {\sc hpm} predict that the end
points from all angles of stress changes (some of the
angles are given at the deflection points) form a
straight line OA. (Instead of a line, a narrow ellipse is
reported in the 2D-simulation of~\cite{hh}. This may be a
result of the fact that the stationarity of $T_g$ is
briefly violated when the stress rate is reversed, during
which the system is rather less plastic.) OA's angle
$\sigma$ in strain space is usually referred to as the
``flow direction," while the direction in stress space,
along which the plastic deformation is largest (with the
strain starting at O and ending at A) is called the
``yield direction" $\phi$. Since they are not equal, the
flow rule is ``non-associated." In Fig~4, the flow
direction $\sigma$, the yield direction $\phi$, and the
maximal plastic strain (the length of OA), are displayed
as functions of $\sigma_s/P$, with $P=0.2$ MPa. Again,
the similarity between both theories is obvious.

We take all this to be a preliminary confirmation for
the basic idea of slowly sheared granular solids being
viscoelastic, and also for {\sc gsh} as the
appropriate hydrodynamic theory. Next, it should be
interesting to use {\sc gsh} for circumstances, in
which $T_g$ is not stationary and the stress rate
possesses a more complicated form than that given by
Eqs~(\ref{hp1},\ref{hpA},\ref{hpB}). These include
especially sudden changes in the direction of the
strain rate~\cite{her}, such as in cyclic loading or
sound propagation.
Also, one needs to understand
whether {\sc gsh} holds at transitions from granular
solid to liquid, from $v_{ij}=0$ to $v_{ij}\not=0$ for
a stationary stress,
${\textstyle\frac\partial{\partial t}}{\sigma
}_{ij}=0$, in phenomena such as shear-banding.

\begin{figure}[t] \label{fig3}
\begin{center}
\includegraphics[scale=.84]{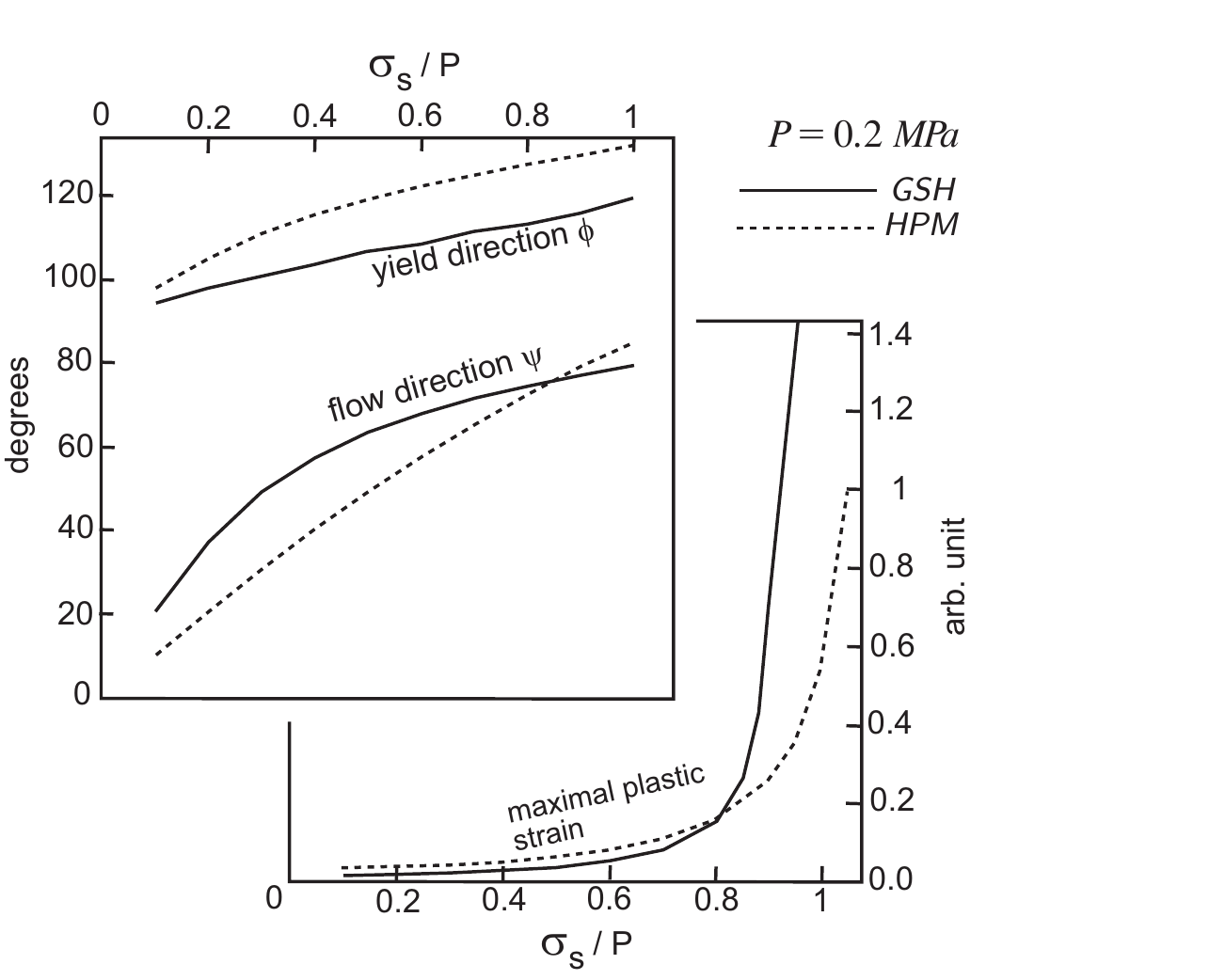}
\end{center}
\caption{Yield direction, flow direction, and the
maximal plastic strain (length of OA), versus
$\sigma_s/P$, for $P=0.2$ MPa, calculated employing
{\sc gsh} and {\sc hpm}, respectively.}
\end{figure}


\begin{thebibliography}{99}
\bibitem{her}A. Niemunis and I. Herle, Mech. of
Cohes.-Frict. Mater., {\bf 2}, 279 (1997). It is
perhaps useful to note that a softening effect may
also be achieved by a $T_g$-dependence of the elastic
coefficient {\cal B}, with $\xi= 5/3$ unchanged.
\end{thebibliography}
\end{document}